\documentclass[12pt,a4paper]{article}
\pdfoutput=1
\usepackage{amsmath} 
\usepackage{amsfonts} 
\usepackage{mathtools} 
\usepackage{amssymb} 
\usepackage{mathrsfs}
\usepackage{subfigure} 
\usepackage{graphicx}
\usepackage{color}

\usepackage{jheppub}

\newcommand{\abs}[1]{|#1|} 
\newcommand{\re}[1]{\text{Re}\left(#1\right)} 
\newcommand{\im}[1]{\text{Im}\left(#1\right)} 
\newcommand{\refEQ}[1]{eq.\,\eqref{#1}} 
\newcommand{\refEQS}[1]{eqs.\,\eqref{#1}} 

\newcommand{\Bd}{B^0_d}
\newcommand{\Bdb}{\bar B^0_d}

\newcommand{\KS}{K_S}
\newcommand{\KL}{K_L}
\newcommand{\KSL}{K_{S,L}}

\newcommand{\KET}[1]{|#1\rangle}
\newcommand{\BRA}[1]{\langle#1|}

\newcommand{\EAmp}[2]{\mathcal A_{#1}^{#2}}

\newcommand{\KETo}{|\Bd\rangle}
\newcommand{\KETot}[1]{|\Bd(#1)\rangle}

\newcommand{\KETob}{|\Bdb\rangle}
\newcommand{\KETobt}[1]{|\Bdb(#1)\rangle}

\newcommand{\KETH}{|B_H\rangle}
\newcommand{\KETL}{|B_L\rangle}

\newcommand{\Lf}{\lambda_{f}}

\newcommand{\Cf}{C_{f}}
\newcommand{\Sf}{S_{f}}
\newcommand{\Rf}{R_{f}}
\newcommand{\Cg}{C_{g}}
\newcommand{\Sg}{S_{g}}
\newcommand{\Rg}{R_{g}}

\newcommand{\C}[1]{C_{#1}}
\renewcommand{\S}[1]{S_{#1}}
\newcommand{\R}[1]{R_{#1}}

\newcommand{\COfont}{\mathscr }
\newcommand{\COC}[3]{\COfont C_{#1}^\om[#2,#3]}
\newcommand{\COS}[3]{\COfont S_{#1}^\om[#2,#3]}
\newcommand{\COCcBASE}{\COfont C_{c}}
\newcommand{\COChBASE}{\COfont C_{h}}
\newcommand{\COScBASE}{\COfont S_{c}}

\newcommand{\COCc}[2]{\COC{c}{#1}{#2}}
\newcommand{\COCh}[2]{\COC{h}{#1}{#2}}
\newcommand{\COSc}[2]{\COS{c}{#1}{#2}}

\newcommand{\FCOCcBASE}{\mathrm C}
\newcommand{\FCOScBASE}{\mathrm S}
\newcommand{\FCOCc}[2]{\FCOCcBASE[#1,#2]}
\newcommand{\FCOSc}[2]{\FCOScBASE[#1,#2]}

\newcommand{\norm}[2]{N_{[#1,#2]}}

\newcommand{\om}{\omega}
\newcommand{\Om}{\Omega}

\graphicspath{{Figs/}}

\title{The signal of ill-defined CPT weakening entanglement in the $\boldsymbol{B_d}$ system}

\author[a]{Jos\'e Bernab\'eu,}
\author[a]{Francisco J. Botella,}
\author[b]{Nick E. Mavromatos,}
\author[c]{Miguel Nebot}
\affiliation[a]{Departament de F\'isica Te\`orica and IFIC, Universitat de Val\`encia - CSIC, E-46100, Spain}
\affiliation[b]{Theoretical Particle Physics and Cosmology Group, Department of Physics, King's College London, Strand, London WC2R 2LS, U.K.}
\affiliation[c]{Centro de F\'\i sica Te\'orica de Part\'\i culas, Instituto Superior T\'ecnico, Universidade de Lisboa, 1049-001 Lisboa, Portugal}
\emailAdd{Jose.Bernabeu@uv.es}
\emailAdd{Francisco.J.Botella@uv.es}
\emailAdd{nikolaos.mavromatos@kcl.ac.uk}
\emailAdd{miguel.r.nebot.gomez@tecnico.ulisboa.pt}
\preprint{{\footnotesize IFIC/16-86, KCL-PH-TH/2016-45, CFTP/16-019}}

\vspace{1.0cm}
\abstract{In the presence of quantum gravity fluctuations (space-time foam),
the CPT operator may be ill-defined. Its perturbative treatment leads 
to a modification of the Einstein-Podolsky-Rosen correlation of the neutral meson system by adding an Entanglement-weakening term of
the wrong exchange symmetry, the $\omega$-effect. In the current paper we
identify how to probe the complex $\omega$ in the entangled $B_d$-system using
Flavour(f)-CP(g) eigenstate decay channels: the connection between
the Intensities for the two time-ordered decays (f, g) and (g, f)
is lost. Appropriate observables are constructed allowing  
independent experimental determinations of Re($\omega$) and Im($\omega$),
disentangled from CPT violation in the evolution Hamiltonian Re($\theta$)
and Im($\theta$). 2-$\sigma$ tensions for both Re($\theta$) and Im($\omega$) are shown to be uncorrelated.

}
\date{\today}
\begin{document}
\maketitle

\section{Introduction\label{SEC:Intro}}
The physics of discrete symmetries in particle and nuclear physics has always been a fascinating subject, since the observation of CP Violation in the neutral Kaon system \cite{Christenson:1964fg}, which was a clear experimental surprise, and set the scene for subsequent precision tests of such discrete symmetries in other systems, including entangled neutral meson factories. Today CP violation in the $K$ and $B_d$ systems, as well as T violation with entangled $B_d$'s \cite{Lees:2012uka}, have been demonstrated experimentally to great accuracy. However, their combination CPT remains unbroken. This is believed to be due to one of the crucial theorems of modern physics, ensuring CPT Invariance of quantum field theory models that are Lorentz invariant, local (in their interactions) and unitary (that is they conserve probability) \cite{Lueders:1957dq}. This is basically a theorem of flat space-time. Quantum gravity or in general deviations from any of the three assumptions may lead to (independent) violations of CPT, which, if observed in nature, would undoubtedly constitute an indication of completely novel physics. Having mentioned quantum gravity, it is worth recalling a corollary by Wald \cite{Wald:1980nm}, according to which a potential decoherence induced during observations in local scattering experiments in which the experimenter  has no access to microscopic quantum gravity degrees of freedom, may lead to an effectively ill-defined CPT quantum mechanical operator.  This observation prompted the authors of \cite{Bernabeu:2003ym} to introduce a different observable for this kind of decoherence-induced CPT violation, termed $\omega$-effect. The $\omega$-effect is different from the situation where CPT violation is violated in the effective hamiltonian, parameterised by the complex $\theta$ parameter. Among other possible sources, $\theta$ can be due to, e.g., Lorentz violation \cite{Kostelecky:1994rn,Colladay:1996iz} as a result of propagation in some Lorentz violating space-time (or otherwise) backgrounds. In the latter case the quantum mechanical operator that implements CPT symmetry is well defined but simply does not commute with the hamiltonian. 
The $\omega$-effect, if observed,  points to an observation of a phenomenon that is exclusively linked to ill-defined nature of the CPT operator, which to date is theoretically  linked only to fundamental decoherence \cite{Wald:1980nm}, independently of any violation of CPT in the hamiltonian. 

Recently, a study for separate direct evidence of T, CP, CPT symmetry violation was accomplished \cite{Bernabeu:2016sgz}. It was based on the precise identification of genuine asymmetry parameters in the time evolution of intensities between the two decays in a B-Factory of entangled neutral $B_d$-meson states. Their values were obtained from the BaBar measurements \cite{Lees:2012uka} of the different Flavour-CP eigenstate decay channels. The concept, put forward in \cite{Banuls:1999aj,Banuls:2000ki}, uses the entangled character of the initial state as the crucial ingredient to (i) connect experimental double  decay rates with specific meson transitions probabilities and (ii) identify the transformed transition to that taken as a reference \cite{Bernabeu:2013qea,Bernabeu:2015hja}. Possible fake effects \cite{Bernabeu:2016sgz} were demonstrated to be well under control by measurements in the same experiment.
The methodology, discussed in \cite{Banuls:1999aj,Banuls:2000ki}, appears to be \cite{Bernabeu:2012ab} crucially dependent on the assumed maximal entanglement between $\Bd$ and $\Bdb$, or between two orthogonal superpositions of them, as given by the Einstein-Podolsky-Rosen (EPR) correlation \cite{Einstein:1935rr} imposed by their decay from the $\Upsilon(4S)$-state with C = -. The corresponding antisymmetric state of the system has two important implications: (i) the program of using Entanglement and the decays as filtering measurements to prepare and detect the meson states can be implemented at any time for the first decay, even in presence of mixing during the previous entangled evolution; (ii) the coefficients of the different time-dependent terms in the double decay rate intensities for the time-ordered decays to $(g,f)$ are related to those for the time-ordered decays to $(f,g)$. The antisymmetry of the entangled state is kept for any two independent states of the neutral mesons, so its evolution leads to a trivial time dependence with definite symmetry under the combined exchange $(f,t_0; g,t_0+t)\to (g,t_0-t;f,t_0)$. As a consequence, the double decay rate intensity (see \refEQ{eq:Intensity:02} below) satisfies for the coefficients of its time dependence with $\om=0$,
\begin{equation}\label{eq:SymRelation:01}
\COChBASE[f,g]=\COChBASE[{g},{f}],\quad \COCcBASE[{f},{g}]=\COCcBASE[g,f]\quad \text{and}\quad \COScBASE[f,g]=-\COScBASE[g,f],
\end{equation}
where the time-ordered decays $(f,g)$ and $(g,f)$ are, in general, not connected by any symmetry transformation. At this level, they can be considered as two different experimental ways of measuring the same quantity when $\om=0$.

In the application to definite Flavour or CP eigenstates decay products, the preparation by maximal entanglement of the initial state of a single neutral meson is usually referred to as ``flavour tagging'' $\Bd$, $\Bdb$, or ``CP tagging'' $B_+$, $B_-$. The underlying assumption considers $\Bd$, $\Bdb$ as two states of the same field, in order to impose Bose statistics with charge conjugation C and permutation $\mathcal P$ with C$\mathcal P$ = +, and it may be invalidated if the CPT operator cannot be intrinsically well defined, as mentioned above. This latter circumstance may occur, for example, in the context of an extended class of quantum gravity models, where the structure of quantum space time at Planckian scales ($10^{-35}$ m) may actually be fuzzy, characterised by a ``foamy'' nature (space-time foam) \cite{Ellis:1992pm,Ellis:1995xd,Bernabeu:2003ym}. Let us emphasize once more that this kind of CPT breaking is different from an explicit CPT violation in the hamiltonian dynamics such that [CPT, H] $\neq 0$, as conventionally introduced, in the context of the Weisskopf-Wigner approach \cite{Weisskopf:1930ps,Lee:1957qq,Enz:1965tr} for the neutral meson system, in the mass matrix. This last CPT violation does not invalidate the analysis followed in \cite{Bernabeu:2016sgz} and, in fact, genuine observables for CPT violation were found with their values obtained from experiment. However, the CPT breaking associated to ``ill-defined'' particle-antiparticle states modifies the EPR correlation, producing the aforementioned $\om$-effect \cite{Bernabeu:2003ym,Alvarez:2004tj,Alvarez:2006ry}. Treating it in perturbation theory, in such a way that we still talk the language of $\Bd$, $\Bdb$, the perturbed two-particle state will contain a component of the ``wrong'' symmetry at the instant of their production by the decay of $\Upsilon(4S)$:
\begin{equation}
\label{eq:InitialState:01}
\KET{\Psi_0}\propto \KETo\KETob-\KETob\KETo+\om\big[\KETo\KETob+\KETob\KETo\big] ,
\end{equation}
where $\om = \abs{\om} e^{i\Om}$ is a complex CPT-breaking parameter \cite{Bernabeu:2003ym,Alvarez:2004tj}, associated with the non-identical particle nature of the neutral meson and antimeson states. The presence of an $\omega$-effect weakens the entanglement of the initial state (\ref{eq:InitialState:01}), 
as follows  from the fact that when $\omega=\pm 1$ the state simply reduces to a product state, whilst when $\omega=0$ the state is fully entangled.

We emphasize that the modification in \refEQ{eq:InitialState:01} is due to the loss of indistinguishability of $\Bd$ and $\Bdb$ and not due to violation of symmetries in the production process. Evidently, the probabilities for the two states connected by a permutation are different due to the presence of $\omega$. This modification of the initial state vector has far-reaching consequences for the concept of meson tagging and for the relation of the time dependent intensities between the decays to time-ordered $(f,g)$ and $(g,f)$ channel\footnote{Another important  aspect of the $\om$ effect is its dynamical generation during a decoherence evolution, in which the particle interacts with its gravitational environment, for instance. As discussed in \cite{Bernabeu:2006av}, a time-dependent contribution to the $\om$ parameter may be generated in specific models of quantum decoherence, which could be present even if the initial state has an $\om$=0. The relative magnitude of $\re{\om(t)}$ and $\im{\om(t)}$ in this case depends crucially on the decoherence space-time foam model used, but their generic form involves oscillatory dependences on time. Specifically, if one ignores conventional CPT violating effects, then the analysis of \cite{Bernabeu:2006av} has shown that  the evolution of an entangled two-particle state  contains in certain quantum space-time-foam models  time-dependent $\omega (t)$ parts, which to leading order in appropriate small quantities assume the form:
\begin{equation}\label{timeomega}
|\psi \rangle \ni e^{-i ( \lambda_0^{(1)} + \lambda_0^{(2)})t} \, \omega (t) \Big(|k, \uparrow \rangle^{(1)} \, |-k, \uparrow \rangle^{(2)}  - |k, \downarrow \rangle^{(1)} \, |-k, \downarrow \rangle^{(2)}\Big)~, \quad \omega (t) \sim \omega_0 \, {\rm sin}(2 |\Delta E| \, t)
\end{equation}
that is purely generated by the evolution with no $\om$ effect in the initial state $t=0$. In the above formula, the superscripts $(i), i=1,2 $ refer to individual particles, $k$ is the momentum of the particle (assuming the decaying initial state to be at rest, for brevity), $\Delta E $ is the energy difference between the appropriate single particle states, and the arrows denote the corresponding quantum numbers of a generic two state system, while $\lambda_0$ are the energy eigenvalues. The parameter $\om_0$ in (\ref{timeomega}) 
is in general complex. In some concrete models of space-time quantum foam it could be purely imaginary~\cite{Bernabeu:2006av}. In the present work we shall consider only constant $\om$ in the initial state (\ref{eq:InitialState:01}). We reserve details for the phenomenology of a time-dependent $\om$-effect, generated during the evolution, for a future publication.}.

In what follows we will study the non-trivial time evolution of \refEQ{eq:InitialState:01}, in the simplified but physically relevant case of a time independent $\om$, in order to (i) establish the appearance of terms of the (previously forbidden) type $\KETo\KETo$ and $\KETob\KETob$, and (ii) introduce a set of observables, which  actually serve as a direct way for measuring $\om$, based on the violation of the relations \refEQ{eq:SymRelation:01}, i.e. using as observables for $\om\neq 0$:
\begin{equation}
\label{eq:NoSymRelation:01}
\COCh{f}{g}-\COCh{g}{f},\quad \COCc{f}{g}-\COCc{g}{f}\quad \text{and}\quad \COSc{f}{g}+\COSc{g}{f},
\end{equation}
and checking experimentally the robustness of the correlation between the two states assumed during the tagging. This paper demonstrates that the comparison between the double decay rate Intensities for time-ordered ($f=$ Flavour, $g=$ CP) eigenstate decay products and $(g,f)$ is sensitive to both $\re{\om}$ and $\im{\om}$.

\section{Time evolution\label{SEC:Evolution}}

\subsection{Double decay rates, time dependent intensities\label{sSEC:Intensity}}
The eigenstates of the effective hamiltonian $\mathbf{H}$ are\footnote{As is commonplace, subindices ``H'' and ``L'' correspond to the heavy and light $B_d$ states.} 
\begin{equation}
\label{eq:H:eigenstates:01}
\begin{split}
&\mathbf{H}\KETH=\mu_H\KETH,\quad \KETH=p_H\KET{\Bd}+q_H\KET{\Bdb},\\
&\mathbf{H}\KETL=\mu_L\KETL,\quad\ \KETL=p_L\KET{\Bd}-q_L\KET{\Bdb}.
\end{split}
\end{equation}
In terms of them
\begin{multline}
\label{eq:InitialState:02}
\KET{\Psi_0}\propto\KETL\KETH-\KETH\KETL\\
+\om\Big\{\theta\big[\KETH\KETL+\KETL\KETH\big]+(1-\theta)\frac{p_L}{p_H}\KETH\KETH-(1+\theta)\frac{p_H}{p_L}\KETL\KETL\Big\},
\end{multline}
where $\theta$ is a CP and CPT violating complex parameter given by $\theta=\frac{\mathbf{H}_{22}-\mathbf{H}_{11}}{\mu_H-\mu_L}$.
 The time evolution of two-meson flavour states is 
\begin{equation}
\label{eq:TwoMesonTimeEvol:01}
\begin{pmatrix}
\KET{\mathrm{A}(t)}\\ \KETot{t}\KETot{t} \\ \KET{\mathrm{S}(t)}\\ \KETobt{t}\KETobt{t}
\end{pmatrix}=
e^{-\Gamma\,t}e^{-i2M\,t}
\begin{pmatrix}
1 & \begin{matrix} 0 & & & 0 & & & 0 \end{matrix}\\
\begin{matrix} 0 \\ 0 \\ 0 \end{matrix} & \mathrm{C} + \mathrm{E}_{[+]} e^{i\Delta\mu\,t}+\mathrm{E}_{[-]} e^{-i\Delta\mu\,t}
\end{pmatrix}
\begin{pmatrix}
\KET{\mathrm{A}}\\ \KETo\KETo \\ \KET{\mathrm{S}}\\ \KETob\KETob
\end{pmatrix},
\end{equation}
where $\mu_H+\mu_L=2M-i\Gamma$, $\mu_H-\mu_L=\Delta\mu=\Delta M-i\frac{\Delta\Gamma}{2}$,
\begin{equation}
\label{eq:TwoMesonTimeEvol:01b}
\begin{matrix}
\KET{\mathrm{A}(t)}=\frac{1}{\sqrt 2}\left[\KETot{t}\KETobt{t}-\KETobt{t}\KETot{t}\right],\\
\KET{\mathrm{S}(t)}=\frac{1}{\sqrt 2}\left[\KETot{t}\KETobt{t}+\KETobt{t}\KETot{t}\right],
\end{matrix}
\end{equation}
and
\begin{equation}
\KETot{t}=e^{-i\mathbf{H}t}\,\KETo,\quad \KETobt{t}=e^{-i\mathbf{H}t}\,\KETob.
\end{equation}
The $\mathrm{C}$, $\mathrm{E}_{[\pm]}$ matrices are
\begin{equation}
\label{eq:TwoMesonTimeEvol:02}
\mathrm{C}=
\begin{pmatrix}
\frac{1}{2}(1-\theta^2) & \frac{1}{\sqrt 2}\frac{q}{p}\theta\sqrt{1-\theta^2} & -\frac{1}{2}\frac{q^2}{p^2}(1-\theta^2)\\
\frac{1}{\sqrt 2}\frac{p}{q}\theta\sqrt{1-\theta^2} & \theta^2 & -\frac{1}{\sqrt 2}\frac{q}{p}\theta\sqrt{1-\theta^2}\\
-\frac{1}{2}\frac{p^2}{q^2}(1-\theta^2) & -\frac{1}{\sqrt 2}\frac{p}{q}\theta\sqrt{1-\theta^2} & \frac{1}{2}(1-\theta^2)
\end{pmatrix},
\end{equation}

\begin{equation}
\label{eq:TwoMesonTimeEvol:03}
\mathrm{E}_{[+]}=
\begin{pmatrix}
\frac{1}{4}(1+\theta)^2 & -\frac{1}{2\sqrt 2}\frac{q}{p}(1+\theta)\sqrt{1-\theta^2} & \frac{1}{4}\frac{q^2}{p^2}(1-\theta^2)\\
-\frac{1}{2\sqrt 2}\frac{p}{q}(1+\theta)\sqrt{1-\theta^2} & \frac{1}{2}(1-\theta^2) & -\frac{1}{2\sqrt 2}\frac{q}{p}(1-\theta)\sqrt{1-\theta^2}\\
\frac{1}{4}\frac{p^2}{q^2}(1-\theta^2) & -\frac{1}{2\sqrt 2}\frac{p}{q}(1-\theta)\sqrt{1-\theta^2} & \frac{1}{4}(1-\theta)^2
\end{pmatrix},
\end{equation}

\begin{equation}
\label{eq:TwoMesonTimeEvol:04}
\mathrm{E}_{[-]}=
\begin{pmatrix}
\frac{1}{4}(1-\theta)^2 & \frac{1}{2\sqrt 2}\frac{q}{p}(1-\theta)\sqrt{1-\theta^2} & \frac{1}{4}\frac{q^2}{p^2}(1-\theta^2)\\
\frac{1}{2\sqrt 2}\frac{p}{q}(1-\theta)\sqrt{1-\theta^2} & \frac{1}{2}(1-\theta^2) & \frac{1}{2\sqrt 2}\frac{q}{p}(1+\theta)\sqrt{1-\theta^2}\\
\frac{1}{4}\frac{p^2}{q^2}(1-\theta^2) & \frac{1}{2\sqrt 2}\frac{p}{q}(1+\theta)\sqrt{1-\theta^2} & \frac{1}{4}(1+\theta)^2
\end{pmatrix}.
\end{equation}
\noindent In \refEQS{eq:TwoMesonTimeEvol:02}-\eqref{eq:TwoMesonTimeEvol:04}, $\frac{q}{p}$ is the usual meson mixing quantity given by $\frac{q^2}{p^2}=\frac{\mathbf{H}_{21}}{\mathbf{H}_{12}}=\frac{q_Hq_L}{p_Hp_L}$. Before addressing actual observables, it is worth noting that, attending to \refEQ{eq:TwoMesonTimeEvol:01}, it is clear that the presence of the symmetric state $\KET{\mathrm{S}}$ in \refEQ{eq:InitialState:01} induces the appearance of $\KETo\KETo$ and $\KETob\KETob$ states.

The transition amplitude for the decay of the first state into $\KET{f}$ at time $t_{0}$, and then the second state into $\KET{g}$ at time $t+t_{0}$ is $\BRA{f,t_0;g,t+t_0}T\KET{\Psi_0}$. Squaring and integrating over $t_{0}$, the double decay rate $I(f,g;t)$ is obtained:
\begin{equation}
\label{eq:Intensity:01}
I(f,g;t)=\int_0^\infty\!\!\!\!\!  dt_0\,\abs{\BRA{f,t_0;g,t+t_0}T\KET{\Psi_0}}^2\,.
\end{equation}
Expanding to first order in $\om$, $\theta$ and taking $\Delta\Gamma=0$, $I(f,g;t)$ has the following form for generic $f$ and $g$ decay channels\footnote{In the notation of reference \cite{Bernabeu:2016sgz}, $\COChBASE[f,g]$, $\COCcBASE[f,g]$ and $\COScBASE[f,g]$ are the $\om\to 0$ limit of (respectively) $\COCh{f}{g}$, $\COCc{f}{g}$ and $\COSc{f}{g}$ in \refEQS{eq:Intensity:02a}-\eqref{eq:Intensity:02c}.}:
\begin{equation}
\label{eq:Intensity:02}
I(f,g;t)=\frac{\langle\Gamma_f\rangle\langle\Gamma_g\rangle}{\Gamma}e^{-\Gamma\,t}
\big\{
\COCh{f}{g}+\COCc{f}{g}\cos(\Delta Mt)+\COSc{f}{g}\sin(\Delta Mt)
\big\},
\end{equation}
with
\begin{multline}
\label{eq:Intensity:02a}
\COCh{f}{g}=\norm{f}{g}\Big[
1-\Rf\Rg+\re{\theta}(\Cg\Rf+\Cf\Rg)-\im{\theta}(\Sf+\Sg)\\
+\frac{1}{1+(x/2)^2}
\big\{
(2\Cf+x\Sf)\re{\om}+(x\Cf-2\Sf)\Rg\im{\om}
\big\}
\Big],
\end{multline}
\begin{multline}
\label{eq:Intensity:02b}
\COCc{f}{g}=\norm{f}{g}\Big[
-(\Cf\Cg+\Sf\Sg)-\re{\theta}(\Cg\Rf+\Cf\Rg)+\im{\theta}(\Sf+\Sg)\\
+\frac{1}{1+(x/2)^2}
\big\{
-(2\Cg+x\Sg)\re{\om}+(-x\Cg+2\Sg)\Rf\im{\om}
\big\}
\Big],
\end{multline}
\begin{multline}
\label{eq:Intensity:02c}
\COSc{f}{g}=\norm{f}{g}\Big[
(\Cg\Sf-\Cf\Sg)+\re{\theta}(\Rg\Sf-\Rf\Sg)+\im{\theta}(\Cf-\Cg)\\
+\frac{1}{1+(x/2)^2}
\big\{
(x\Cg-2\Sg)\re{\om}-(2\Cg+x\Sg)\Rf\im{\om}
\big\}
\Big],
\end{multline}
where, in terms of the decays amplitudes $\BRA{f}T\KET{\Bdb}\equiv\bar A_f$ and $\BRA{f}T\KET{\Bd}\equiv A_f$, the following parameters are used\footnote{By construction $\Cf^2+\Sf^2+\Rf^2=1$.}:
\begin{equation}
\Lf\equiv \frac{q}{p}\frac{\bar A_f}{A_f},\quad \Cf=\frac{1-\abs{\Lf}^2}{1+\abs{\Lf}^2},\quad \Rf= \frac{2\re{\Lf}}{1+\abs{\Lf}^2},\quad \Sf= \frac{2\im{\Lf}}{1+\abs{\Lf}^2},
\end{equation}
\begin{equation}
\norm{f}{g}=\frac{1-\delta^2}{(1+\abs{\om}^2)(1-\delta\Cf)(1-\delta\Cg)},\ \text{and }\langle\Gamma_f\rangle=\frac{\abs{\bar A_f}^2+\abs{A_f}^2}{2}.
\end{equation}
In addition, $x=\frac{\Delta M}{\Gamma}\simeq 0.77$ and $\delta=\frac{1-\abs{q/p}^2}{1+\abs{q/p}^2}\simeq 1-2\times 10^{-3}$. 
It is worth reminding that for flavour-specific decay channels $X+\ell^\pm$ (``$\ell^\pm$'' for short in the following), we have $\C{\ell^\pm}=\pm 1$, $\R{\ell^\pm}=\S{\ell^\pm}=0$.

\subsection{Sensitivity to ${\omega}$\label{sSEC:Sensitivity}}

Coming back to the transition amplitude $\BRA{f,t_0;g,t+t_0}T\KET{\Psi_0}$, it has the following structure:
\begin{equation}\label{eq:TransAmp:01}
\BRA{f,t_0;g,t+t_0}T\KET{\Psi_0}\propto e^{(-iM-\Gamma/2)(2t_0+t)}
\begin{pmatrix}
[e^{-i\Delta\mu\,t/2}\EAmp{f}{L}\EAmp{g}{H}-e^{i\Delta\mu\,t/2}\EAmp{f}{H}\EAmp{g}{L}]\\
+\om\theta[e^{-i\Delta\mu\,t/2}\EAmp{f}{L}\EAmp{g}{H}+e^{i\Delta\mu\,t/2}\EAmp{f}{H}\EAmp{g}{L}]\\
+\om(1-\theta)\frac{p_L}{p_H}e^{-i\Delta\mu\,(t_0+t/2)}\EAmp{f}{H}\EAmp{g}{H}\\
-\om(1+\theta)\frac{p_H}{p_L}e^{i\Delta\mu\,(t_0+t/2)}\EAmp{f}{L}\EAmp{g}{L}
\end{pmatrix}.
\end{equation}
The prefactor $e^{(-iM-\Gamma/2)(2t_0+t)}$ gives a global $e^{-2\Gamma t_0}e^{-\Gamma t}$ dependence in $\abs{\BRA{f,t_0;g,t+t_0}T\KET{\Psi_0}}^2$. One can readily observe that the $\om$-dependent terms, even for $\theta=0$ (i.e. already for the leading $\om$ contribution), do introduce an additional non-trivial $t_0$ dependence. Ignoring that $e^{(-iM-\Gamma/2)(2t_0+t)}$ prefactor, it is clear that combining the transformations $t\mapsto -t$ and $f\leftrightarrows g$, the first contribution, the standard $\om=0$ one, just receives a $(-)$ sign. This implies that, in the absence of $\om$, in the $t$-dependence of $I(f,g;t)$,
\begin{equation}
I(f,g;t)\sim e^{-\Gamma t}\left(\COChBASE[f,g]+\COCcBASE[f,g]\cos(\Delta Mt)+\COScBASE[f,g]\sin(\Delta Mt)\right)
\end{equation}
we necessarily have \cite{Bernabeu:2016sgz}: $\COChBASE[f,g]=\COChBASE[g,f]$, $\COCcBASE[f,g]=\COCcBASE[g,f]$ and $\COScBASE[f,g]=-\COScBASE[g,f]$. 

In the presence of $\om\neq 0$ the situation changes drastically. From the remaining contributions in \refEQ{eq:TransAmp:01}, the ones induced by the evolution of the $\om$-dependent term in \refEQ{eq:InitialState:01}, the situation is more involved: the first one, proportional to $\om\theta$ and $t_0$-independent, is clearly invariant under the combination of $f\leftrightarrows g$ and $t\mapsto -t$. The last two terms are separately invariant under $f\leftrightarrows g$, but have no well defined transformation under $t\mapsto -t$; moreover, contrary to the previous contributions, they depend on $t_0$, the time elapsed between production of the $B\bar B$ pair and the first decay\footnote{For small $\om$ and $\theta$, these terms give the leading $\om$ contributions: the $t_0$ dependence integrated over in \refEQ{eq:Intensity:01} produces extra dilution factors $x/(1+(x/2)^2)^{-1}$ and $1/(1+(x/2)^2)^{-1}$ in \refEQS{eq:Intensity:02a} to \eqref{eq:Intensity:02c}; fortunately, in the $B_d$ system, they do not thwart significantly the sensitivity to $\om$.}. Out of those properties, the simple assignment of symmetry/antisymmetry under $f\leftrightarrows g$ to the $t$-even/$t$-odd terms in $e^{\Gamma\,t}\,I(f,g;t)$, possible when $\om=0$, does not apply when $\om\neq 0$.
This simple remark provides the first understanding of the potential sensitivity to the presence of $\om\neq 0$: while in the absence of $\om$, the measurement of intensities for decays into $f$ and $g$ with the two different orderings (i) first $f$ then $g$ and (ii) first $g$ then $f$, provides two experimentally independent measurements of the same theoretical quantities, in the presence of $\om$ the situation has changed. Deviations from the standard $f\leftrightarrows g$ symmetry properties are a gateway to probe for $\om$. 

The BaBar collaboration performed separate analyses~\cite{Lees:2012uka} for the two different time orderings of the two $B$ meson decays. Previous studies, like \cite{Alvarez:2006ry}, exploited the use of two flavour specific decay channels to obtain bounds on $\re{\om}$ through the appearance of $\KETo\KETo$ and $\KETob\KETob$ states for $t=0$. Equation \eqref{eq:Intensity:02} shows that, using flavour specific channels alone, there is no sensitivity to $\im{\om}$: since $\R{\ell^\pm}=0$, the terms in $\im{\om}$ would be absent\footnote{Equation \eqref{eq:Intensity:02} gives the intensity $I(f,g;t)$ expanded up to linear order in $\theta$ and $\om$: the full result has indeed contributions that depend on $\im{\om}$ and do not vanish when both $f$ and $g$ are flavour specific, but they have additional $\om$ and/or $\theta$ suppressions which make them irrelevant. In any case, the actual fits in section \ref{SEC:Results} are conducted using the full expressions.}. Fortunately enough, besides addressing the two different time orderings, in \cite{Lees:2012uka}, one decay is flavour specific (labelled $\ell^\pm$), while the other is CP specific (decays into $J/\Psi \KSL$, labelled $\KSL$ for short): sensitivity to both $\re{\om}$ and $\im{\om}$ is thus expected.

\subsection{Experimental observables\label{sSEC:Observables}}
In order to reduce experimental uncertainties in the different channels, the BaBar collaboration, in reference \cite{Lees:2012uka}, fixed the constant term and measured the coefficients $\FCOCc{f}{g}$ and $\FCOSc{f}{g}$ of the decay intensity
\begin{equation}
\label{eq:Babar:DoubleRates:01}
\mathbf{g}_{f,g}(t)\propto e^{-\Gamma\, t}\left\{1+\FCOCc{f}{g}\cos(\Delta M\,t)+\FCOSc{f}{g}\sin(\Delta M\,t)\right\},
\end{equation}
using for the $f$ and $g$ states one flavour specific channel, $X\ell^+\nu$ or $X\ell^-\bar\nu$, and one CP eigenstate, $J/\Psi \KS$ or $J/\Psi \KL$.
Obviously we should have
\begin{equation}
\label{eq:Babar:Coefs:01}
\FCOCc{f}{g}=\frac{\COCc{f}{g}}{\COCh{f}{g}}\quad\text{and}\quad \FCOSc{f}{g}=\frac{\COSc{f}{g}}{\COCh{f}{g}},
\end{equation}
where one should remember that in the coefficients $\FCOCc{f}{g}$ and $\FCOSc{f}{g}$, the ordering of $f$ and $g$ means that $f$ corresponds to the first (in time) decay product of the entangled state evolved in time, and $g$ corresponds to the second (in time) decay product. In the case under consideration, the flavour specific decays simplify significantly the expressions, which are, at linear order in $\theta$, $\om$,
\begin{multline}
\FCOCc{\ell^\pm}{g}=
\mp\Cg+\re{\theta}\Rg(\Cg\mp 1)+\im{\theta}\Sg(1\mp\Cg)\\
+\frac{1}{1+(x/2)^2}\left\{
-x\Sg\re{\om}+x\Cg\Rg\im{\om}
\right\},
\end{multline}
\begin{multline}
\FCOSc{\ell^\pm}{g}=
\mp\Sg+\re{\theta}\Sg\Rg+\im{\theta}(\pm 1-\Cg\mp\Sg^2)\\
+\frac{1}{1+(x/2)^2}\left\{
x\Cg\re{\om}+x\Sg\Rg\im{\om}
\right\}.
\end{multline}
In the presence of $\om$, the time ordering definite symmetry is not valid anymore and therefore it is relevant to write the completely different coefficients
\begin{multline}\label{obs3}
\FCOCc{f}{\ell^\pm}=
\mp\Cf+\re{\theta}\Rf(\Cf\mp 1)+\im{\theta}\Sf(1\mp\Cf)\\
+\frac{1}{1+(x/2)^2}\left\{
\pm (2(\Cf^2-1)+x\Cf\Sf))\re{\om}\mp x\Rf\im{\om}
\right\},
\end{multline}
\begin{multline}\label{obs4}
\FCOSc{f}{\ell^\pm}=
\pm\Sf-\re{\theta}\Sf\Rf+\im{\theta}(\mp 1+\Cf\pm\Sf^2)\\
+\frac{1}{1+(x/2)^2}\left\{
\pm(x(1-\Sf^2)-2\Cf\Sf)\re{\om}\mp 2\Rf\im{\om}
\right\}.
\end{multline}
As anticipated, $\FCOCc{\ell^\pm}{g}-\FCOCc{g}{\ell^\pm}$ and $\FCOSc{\ell^\pm}{g}+\FCOSc{g}{\ell^\pm}$ are linear in $\om$, and thus the fact that the BaBar collaboration distinguished the different decay time orderings in \cite{Lees:2012uka}, now reveals crucial to disentangle the $\om$ effect:
\begin{multline}\label{obs1}
\FCOCc{\ell^\pm}{g}-\FCOCc{g}{\ell^\pm}=
\frac{1}{1+(x/2)^2}\\ \times \left\{
\left[x\Sg\mp 2(\Cg^2-1)\mp x\Cg\Sg\right]\re{\om}+x\Rg\left[\Cg\pm 1\right]\im{\om}
\right\},
\end{multline}
\begin{multline}\label{obs2}
\FCOSc{\ell^\pm}{g}+\FCOSc{g}{\ell^\pm}=
\frac{1}{1+(x/2)^2}\\ \times \left\{
\left[x\Cg\pm x(1-\Sg^2)\mp 2\Cg\Sg\right]\re{\om}+\Rg\left[x\Sg\mp 2\right]\im{\om}
\right\}.
\end{multline}
These combinations are linearly sensitive not only to $\re{\om}$ but also to $\im{\om}$ when $\Rg\neq 0$. The sensitivity to $\im{\om}$ depends critically on the use of a CP eigenstate channel with large $\Rg$, as is the case with $J/\Psi \KS$ and $J/\Psi \KL$.

\section{Results\label{SEC:Results}}
We are now ready to present the results obtained from a global fit to available BaBar experimental data, following the same statistical treatment as in reference \cite{Bernabeu:2016sgz}. We use the sixteen experimental observables measured by BaBar in \cite{Lees:2012uka}: $\FCOCc{\ell^\pm}{\KSL}$, $\FCOCc{\KSL}{\ell^\pm}$, $\FCOSc{\ell^\pm}{\KSL}$ and $\FCOSc{\KSL}{\ell^\pm}$. Taking into account full covariance information on statistical and systematic uncertainties, we perform a fit in terms of the set of parameters $\{\re{\theta}$, $\im{\theta}$, $\re{\om}$, $\im{\om}$, $\C{\KS}$, $\S{\KS}$,  $\R{\KS}$, $\C{\KL}$, $\S{\KL}$,  $\R{\KL}\}$ with the known constraints $\Cf^2+\Sf^2+\Rf^2=1$. Therefore we generalize the corresponding fit presented in reference \cite{Bernabeu:2016sgz} to the actual situation where deviations from EPR entanglement are present due to the $\omega$-effect \cite{Bernabeu:2003ym}. A more restricted fit is also done in the case where no wrong sign flavour decays are allowed in the $B_d\to J/\Psi K$ decays, that is with $\lambda_{\KS}+\lambda_{\KL}=0$.\newline

\noindent In table \ref{TAB:fit}(I) we present the general result of the fit, whose most salient features are the following:
\begin{itemize}
\item Experimental data -- more precisely the BaBar measurements in \cite{Lees:2012uka} -- are sensitive for the first time to $\im{\om}$, revealing a tantalizing $2.4\sigma$ deviation from $\im{\om}=0$. These observables are also sensitive to $\re{\om}$, but they do not show any significant deviation from $\re{\om}=0$, and the previous determination $\re{\om}=(0.8\pm 4.6)\times 10^{-3}$ \cite{Alvarez:2006ry} -- using semileptonic channels -- is still better than the present one.
\item The results of the fit for the CPT violating parameter $\theta$ -- in the evolution hamiltonian -- are compatible with the previous determination in \cite{Bernabeu:2016sgz} and the one performed by the BaBar collaboration in reference \cite{BABAR:2016tjc}. An exciting $2\sigma$ effect in $\re{\theta}$ is still present.
\item The parameters that measure the presence of wrong flavour decays in $B_d\to J/\Psi K$, i.e. $\C{\KS}-\C{\KL}$, $\S{\KS}+\S{\KL}$ and $\R{\KS}+\R{\KL}$, do not show any significant deviation from zero and the results are consistent with \cite{Bernabeu:2016sgz}.
\item In the case of $\S{\KS}$ and $\R{\KS}$ we observe that they differ by more than $1\sigma$ with respect to the determination in \cite{Bernabeu:2016sgz} without including the $\omega$ effect. Should this persist in the future, it could affect the precise determination of the unitarity triangle angle $\beta$.
\end{itemize}

\noindent In table \ref{TAB:fit}(II) we present the results of the same fit with the additional requirement of not having wrong flavour decays, $\lambda_{\KS}+\lambda_{\KL}=0$. No significant differences were noticed with respect to the conclusions discussed above for the general case. For completeness we show, when relevant, both analyses together in the same plots without further comments.

\begin{table}[h]
\begin{center}
\begin{tabular}{|l|c||l|c|}
\cline{1-4}
\multicolumn{4}{|c|}{(I) Parameters -- General analysis}\\ \hline
$\re{\theta}$ & $\pm(6.11\pm 3.45)10^{-2}$ & $\im{\theta}$ & $(0.99\pm 1.98)10^{-2}$ \\ \hline
$\re{\om}$ & $(1.09\pm 1.60)10^{-2}$ & $\im{\om}$ & $\pm(6.40\pm 2.80)10^{-2}$ \\ \hline
$\S{\KS}$ & $-0.624\pm 0.030$ & $\R{\KS}$ & $\pm(0.781\pm 0.024) $\\ \hline
$\C{\KS}$ & $\left(-1.44\pm 3.28\right) 10^{-2}$\\ \hline
$\S{\KS}+\S{\KL}$ & $(3.7\pm 4.9)10^{-2}$ & $\R{\KS}+\R{\KL}$ & $(-3.27\pm 4.3)10^{-2}$\\ \hline
$\C{\KS}-\C{\KL}$ & $(-6.8\pm 6.3)10^{-2}$ 
\\\hline
\multicolumn{4}{|c|}{(II) Parameters -- $\lambda_{\KS}+\lambda_{\KL}=0$ analysis}\\ \hline
$\re{\theta}$ & $\pm (3.10\pm 1.51)10^{-2}$ & $\im{\theta}$ & $(0.14\pm 1.67)10^{-2}$ \\ \hline
$\re{\om}$ & $ (1.17\pm 1.59)10^{-2}$ & $\im{\om}$ & $\pm(5.46\pm 2.70)10^{-2}$ \\ \hline
$\S{\KS}$ & $-0.640\pm 0.025$ & $\R{\KS}$ & $\pm(0.769\pm 0.022) $\\ \hline
$\C{\KS}$ & $(1.61\pm 1.88) 10^{-2}$\\ \cline{1-2}
\end{tabular}
\caption{Summary of results.\label{TAB:fit}}
\end{center}
\end{table}

\begin{figure}[h!]
\begin{center}
\includegraphics[width=0.4\textwidth]{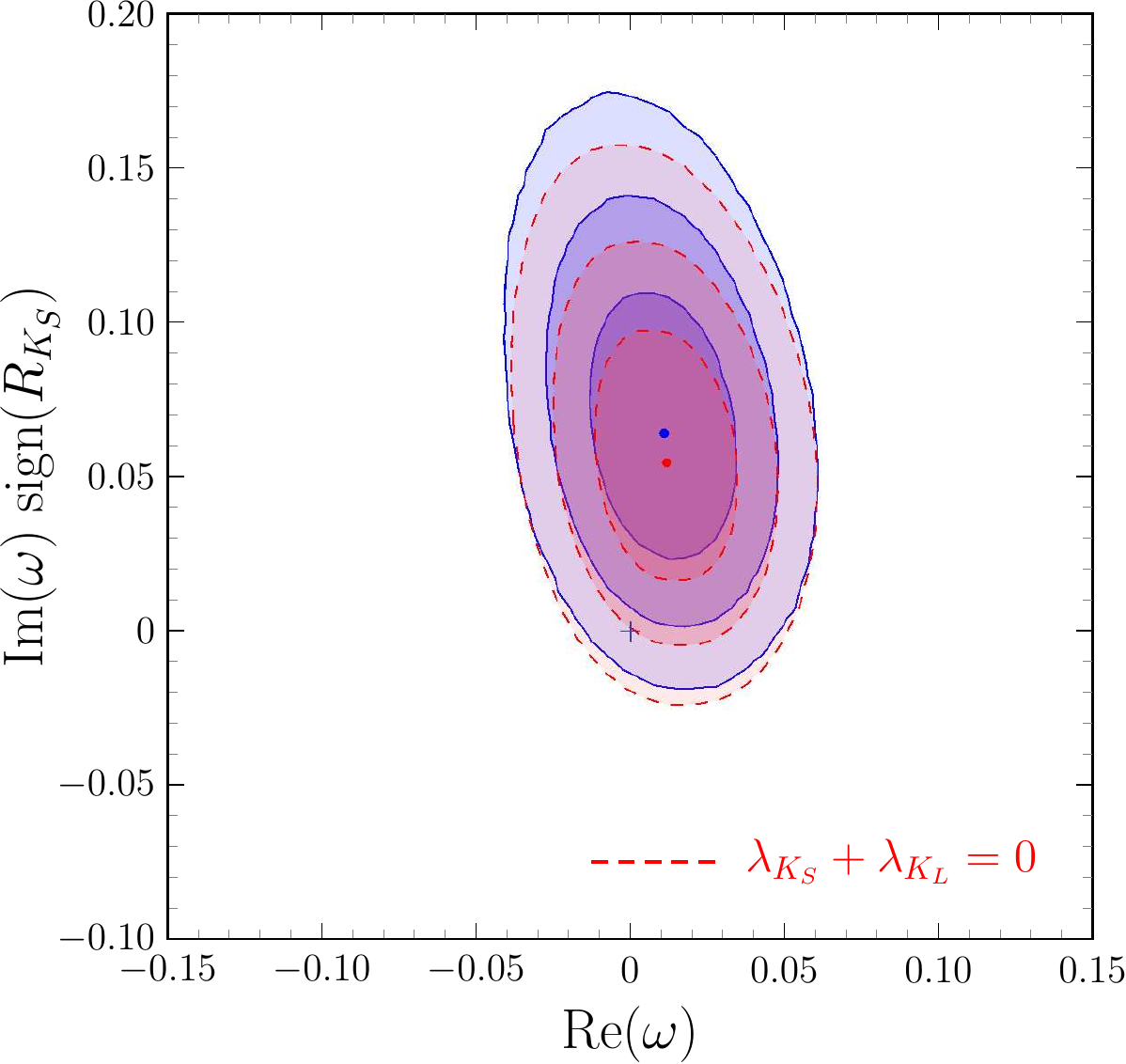}
\caption{$\im{\om}$ vs. $\re{\om}$ in the general fit (blue regions with solid contours), and in the fit with $\lambda_{\KS}+\lambda_{\KL}=0$ (red regions with dashed contours); darker to lighter regions correspond to two-dimensional 68\%, 95\% and 99\% CL. Figures \ref{fig:Omega:04} and \ref{fig:Omega:05} obey the same colour coding for the two fits and the CL regions.\label{fig:Omega:01}}
\end{center}
\end{figure}

\noindent In figure \ref{fig:Omega:01} is shown the result for the new parameters not previously considered in the analyses where EPR entangled initial states where assumed. A deviation of the complex number $\om$ from zero is found at 95\% confidence level. This deviation comes essentially from $\im{\om}$ and it represents a measurement of this parameter for the first time; the measurement of $\re{\om}$ does not improve on the value obtained previously \cite{Alvarez:2006ry} from flavour specific decays.

\begin{figure}[h!]
\begin{center}
\subfigure[Complete fit.\label{fig:Omega:02a}]{\includegraphics[width=0.4\textwidth]{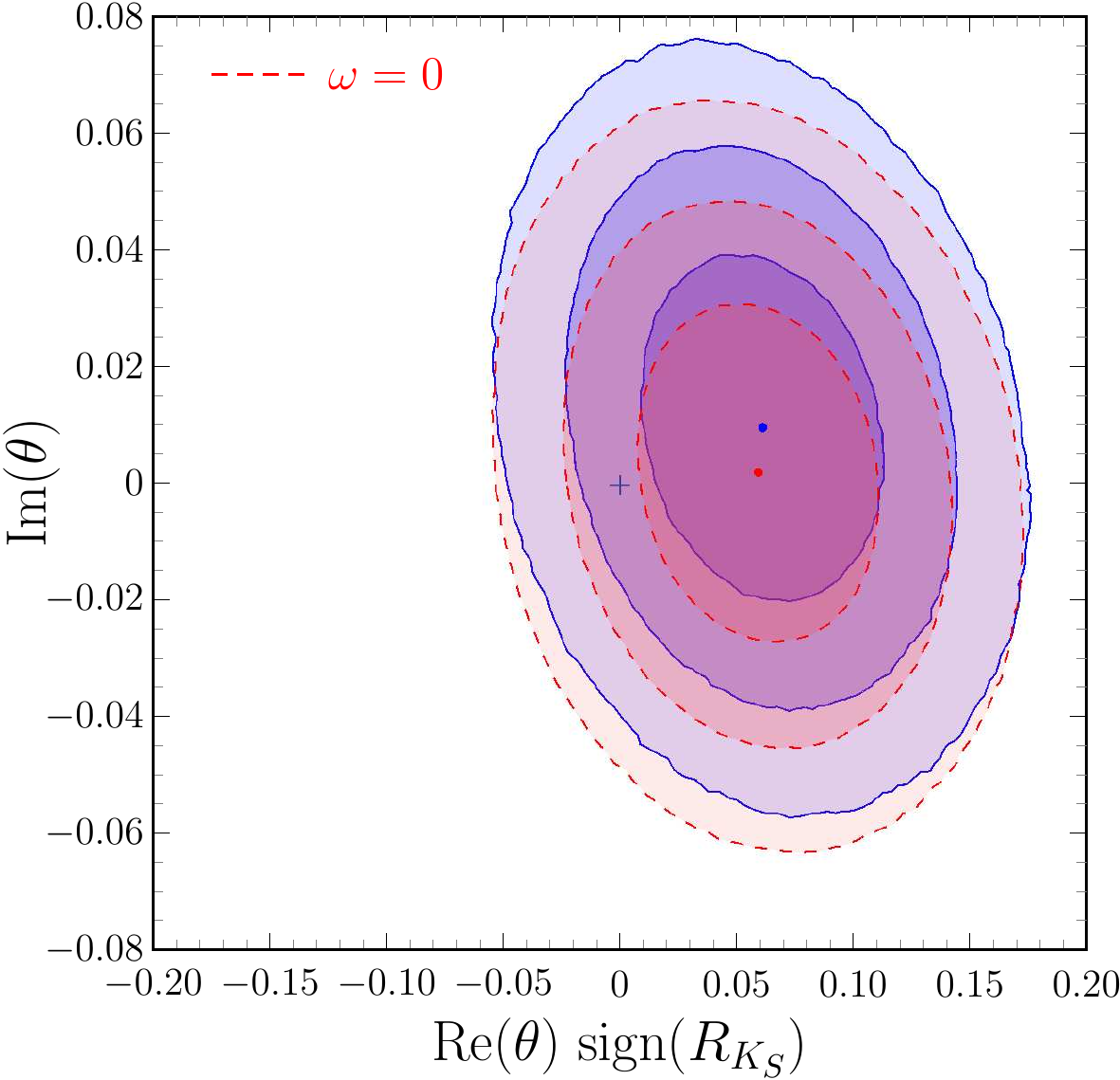}}\qquad
\subfigure[Restricted.\label{fig:Omega:02b}]{\includegraphics[width=0.4\textwidth]{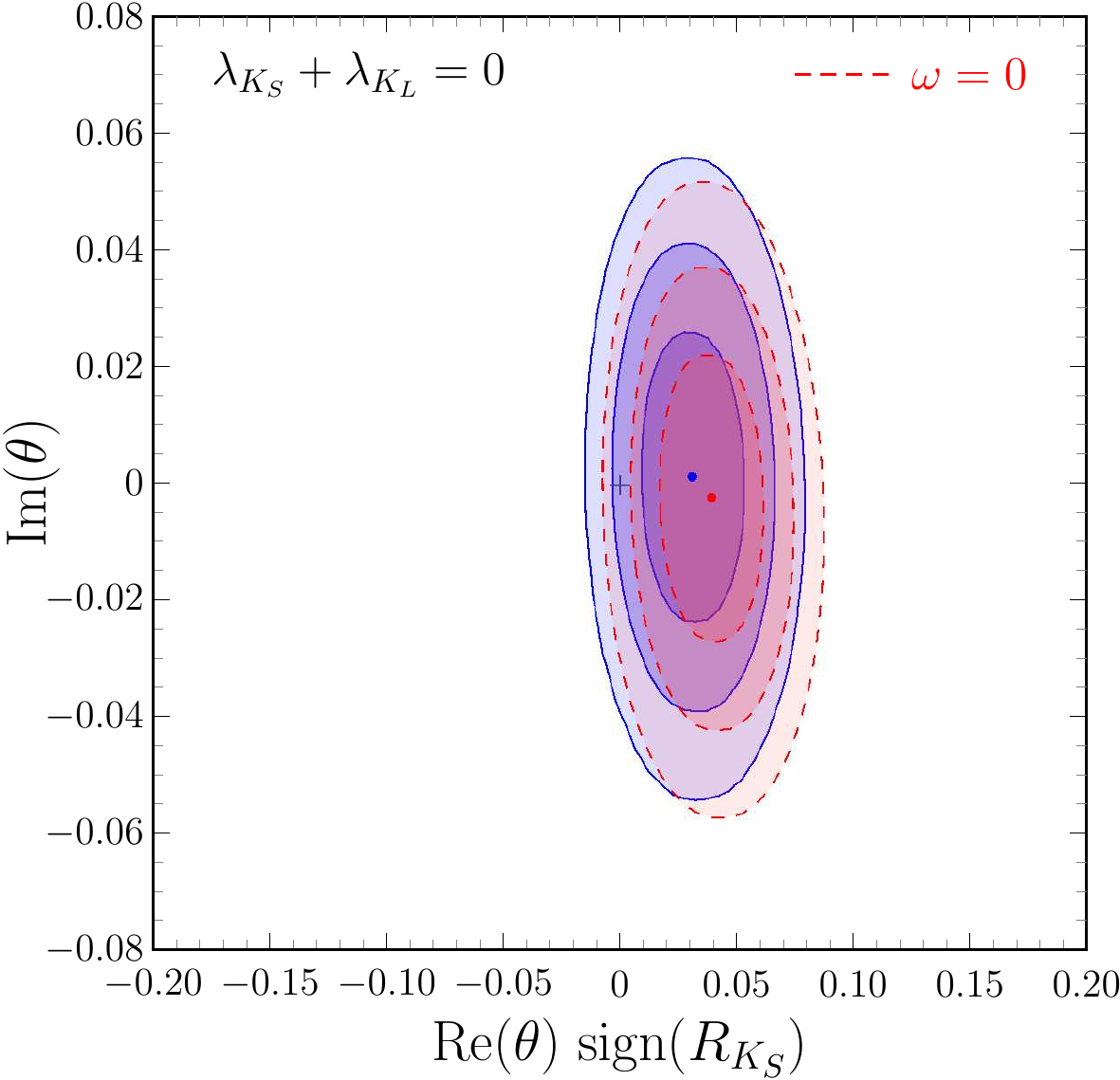}}\\
\caption{Comparison of the results with and without $\om$; $\im{\theta}$ vs. $\re{\theta}$ regions are shown in different scenarios: blue regions with solid contours correspond to fits including the $\om$-effect while red regions with dashed contours correspond fits without $\om$, i.e. with $\om=0$. Panel \ref{fig:Omega:02a} shows the results for the general analyses while panel \ref{fig:Omega:02b} shows the results for the analyses with $\lambda_{\KS}+\lambda_{\KL}=0$.\label{fig:Omega:02}}
\end{center}
\end{figure}

\begin{figure}[h!]
\begin{center}
\subfigure[$\re{\om}$ vs. $\re{\theta}$.\label{fig:Omega:04a}]{\includegraphics[width=0.4\textwidth]{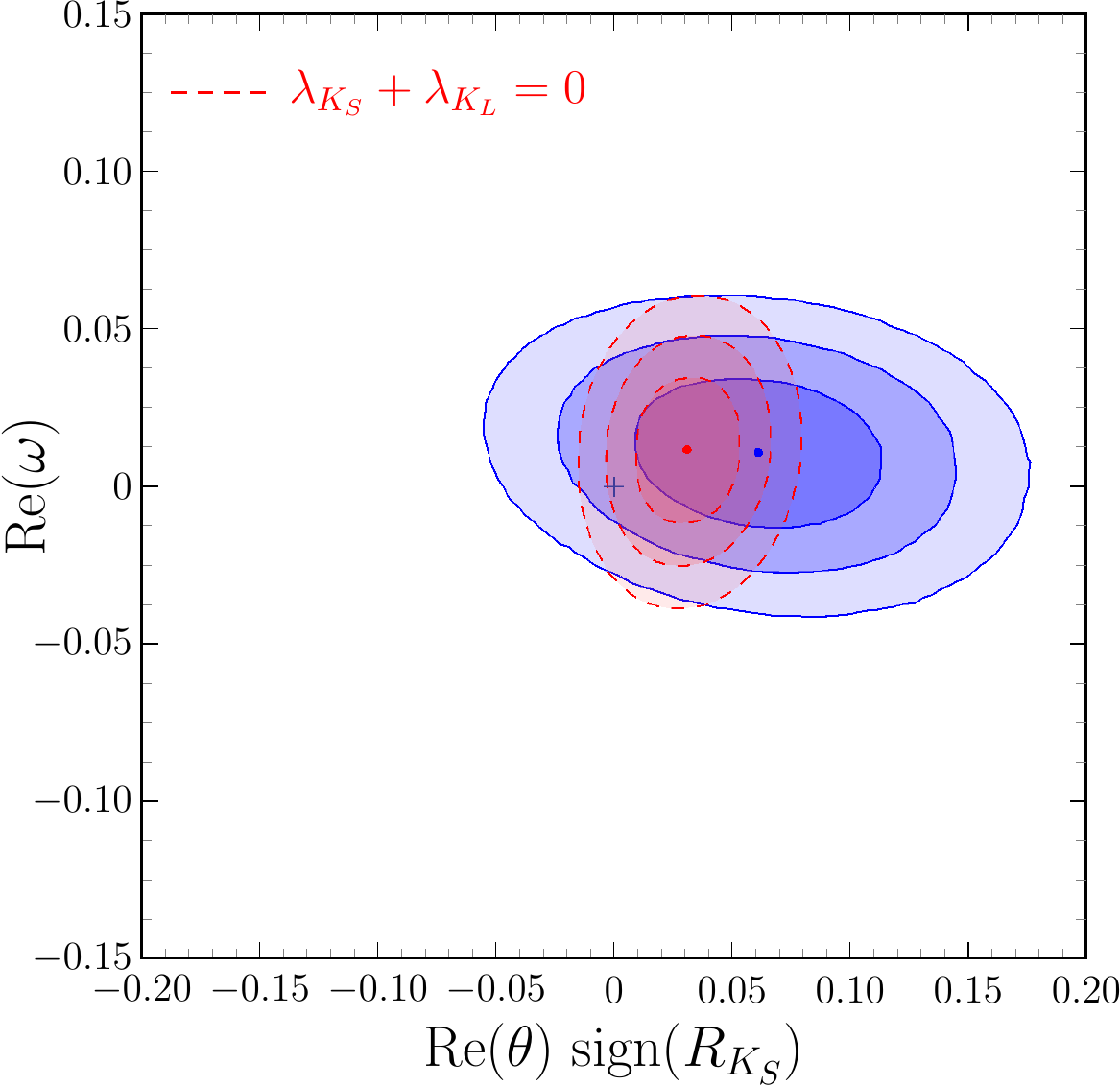}}\qquad
\subfigure[$\re{\om}$ vs. $\im{\theta}$.\label{fig:Omega:04b}]{\includegraphics[width=0.4\textwidth]{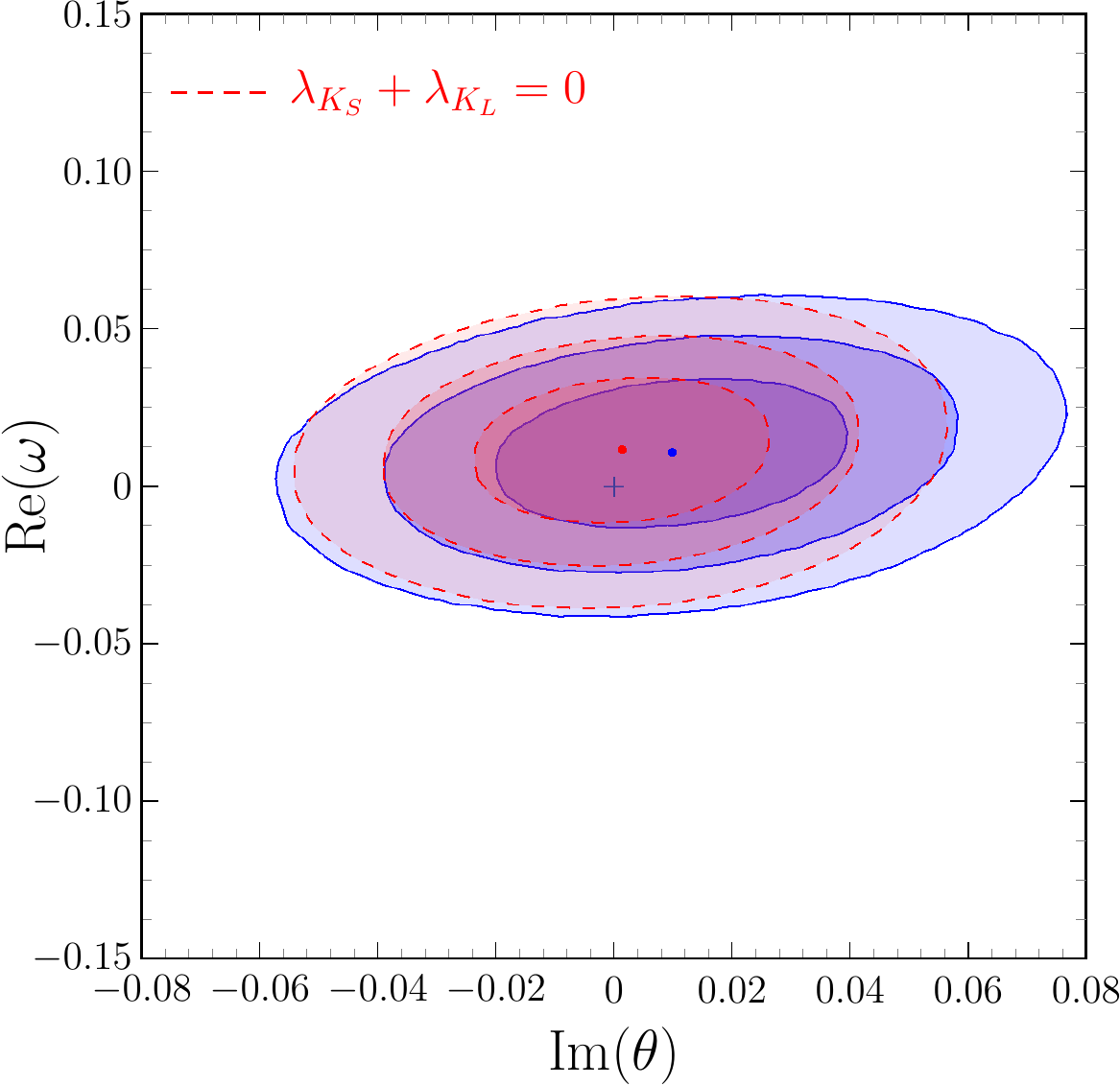}}\\
\subfigure[$\im{\om}$ vs. $\re{\theta}$.\label{fig:Omega:04c}]{\includegraphics[width=0.4\textwidth]{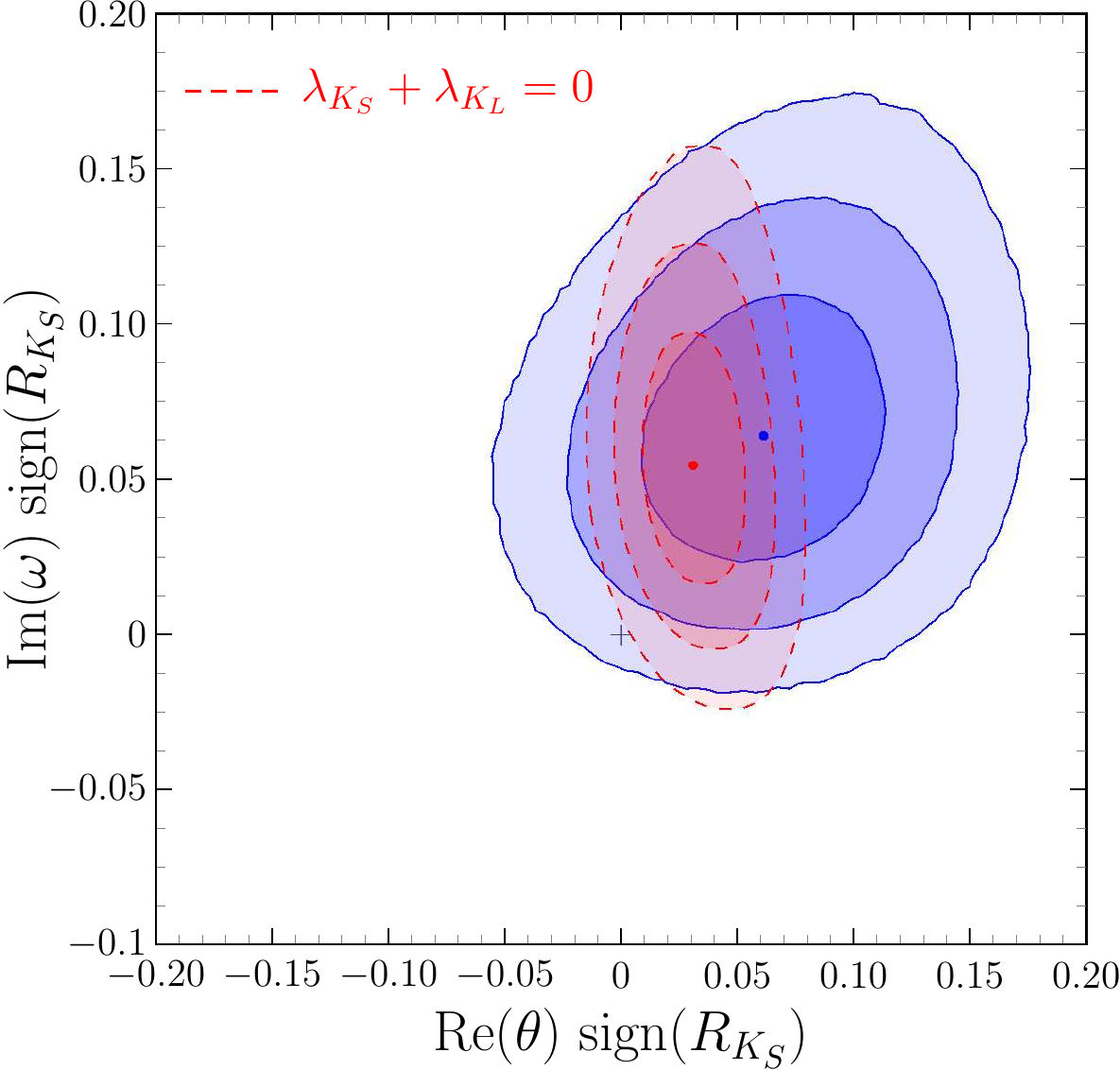}}\qquad
\subfigure[$\im{\om}$ vs. $\im{\theta}$.\label{fig:Omega:04d}]{\includegraphics[width=0.4\textwidth]{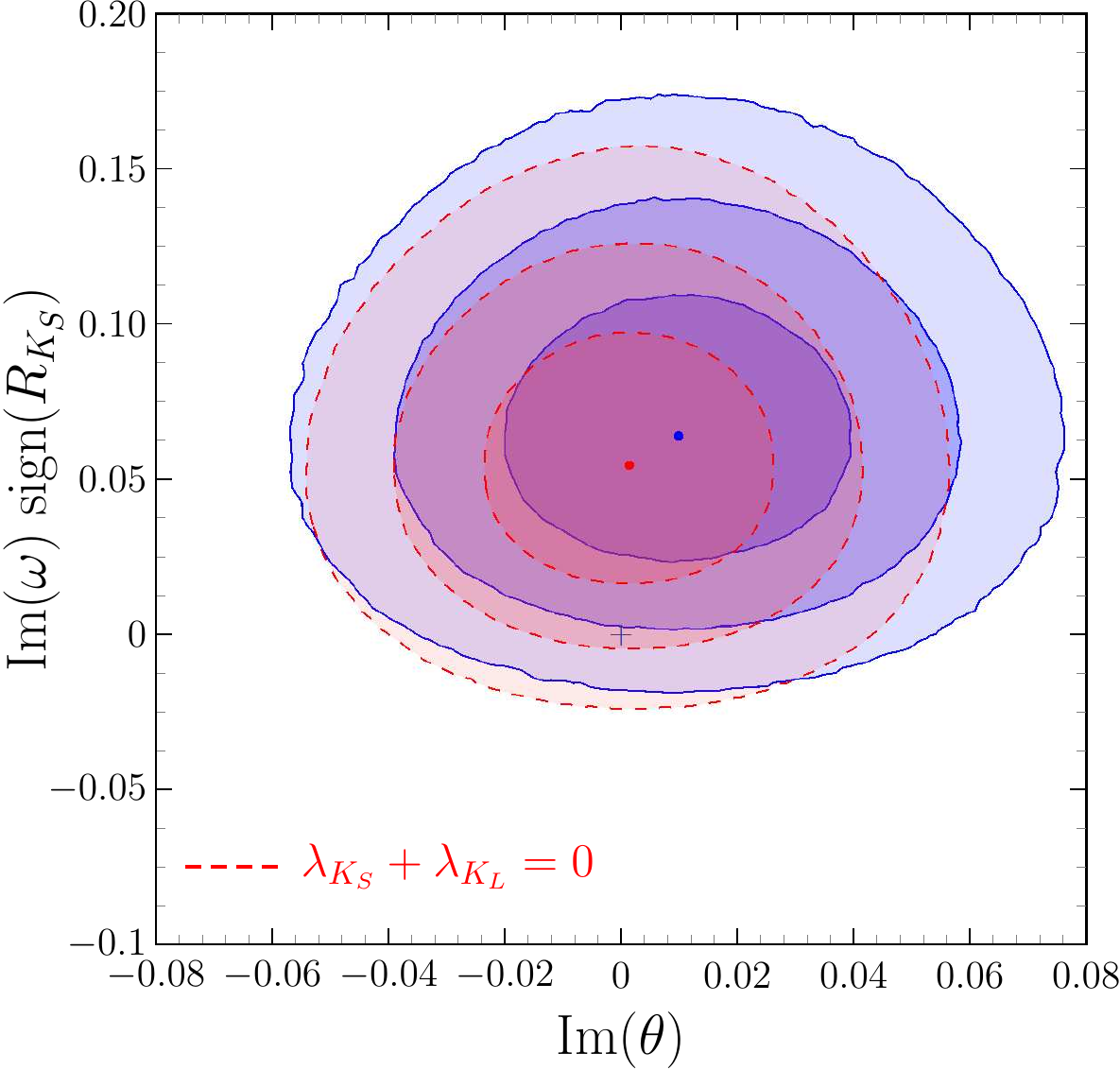}}\\
\caption{Correlations among $\om$ and $\theta$. (Colour coding as indicated in figure \ref{fig:Omega:01}.)\label{fig:Omega:04}}
\end{center}
\end{figure}

\noindent The stability of the fitted value of the complex CPT violating parameter $\theta$ is shown in figures \ref{fig:Omega:02a} and \ref{fig:Omega:02b}, where it is clear that the results for $\re{\theta}$ and $\im{\theta}$ do not change from the constrained case $\om=0$ to the general case with arbitrary $\om$. 
 Cross correlations among the different components of $\theta$ and $\om$ are shown in figure \ref{fig:Omega:04}. For example, figure \ref{fig:Omega:04c} shows the independence of $\im{\om}$ and $\re{\theta}$: furthermore one can see in that figure that the point $(0,0)$ in this projection is at more than $2.5\sigma$ from the best fit values (or even at $3\sigma$ in the $\lambda_{\KS}+\lambda_{\KL}=0$ constrained analysis).

\begin{figure}[h!]
\begin{center}
\subfigure[$\R{\KS}$ vs. $\im{\om}\text{sign}(\R{\KS})$.\label{fig:Omega:05a}]{\includegraphics[width=0.45\textwidth]{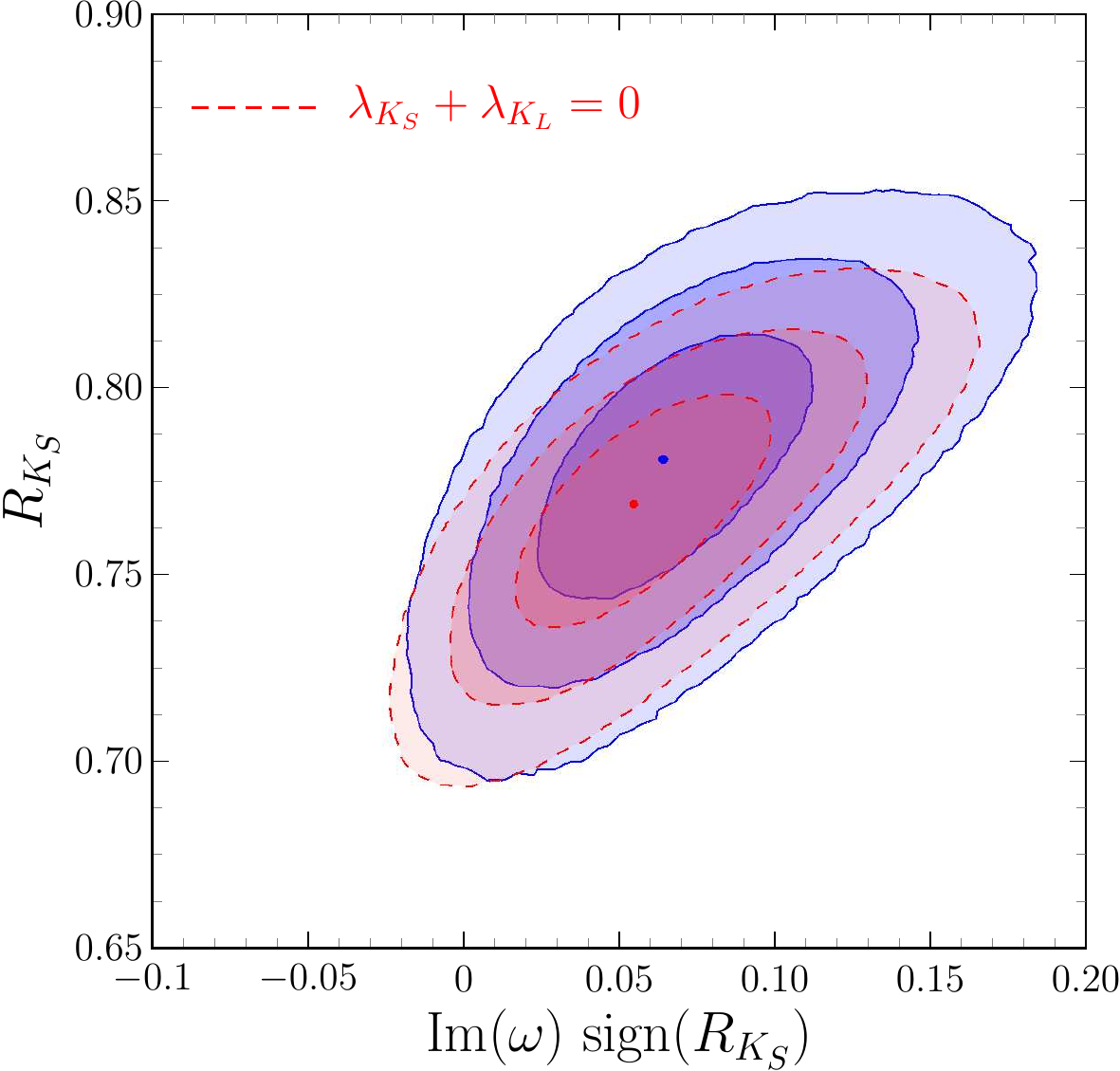}}\qquad
\subfigure[$\S{\KS}$ vs. $\im{\om}\text{sign}(\R{\KS})$.\label{fig:Omega:05b}]{\includegraphics[width=0.46\textwidth]{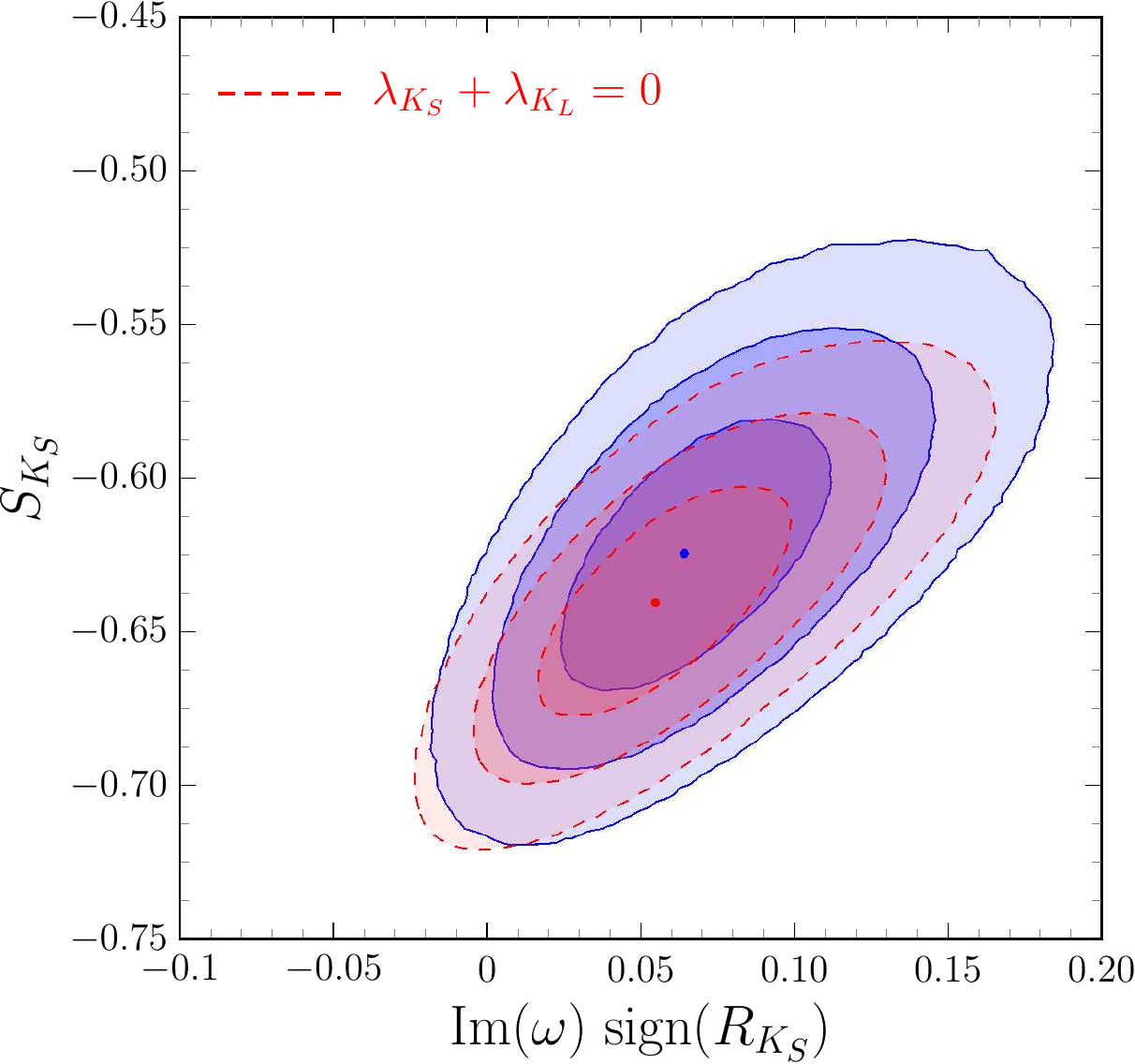}}\\
\caption{$\R{\KS}$ and $\S{\KS}$ vs. $\im{\om}$; the presence of $\om$ affects the extracted values of $\R{\KS}$ and $\S{\KS}$. (Colour coding as indicated in figure \ref{fig:Omega:01}.)\label{fig:Omega:05}}
\end{center}
\end{figure}

\noindent  Finally, in figure \ref{fig:Omega:05}, one can see the near linear correlation among $\im{\om}$ and $\R{\KS}$. This explains why the presence of $\om$ affects both $\R{\KS}$ and $\S{\KS}$.

\clearpage
\section{Conclusions\label{SEC:Conc}}
In the present article we have discussed the possibility of probing the entanglement-weakening CPT Violating parameter $\omega$, that potentially signifies the breakdown of CPT operation as a result of quantum decoherence of matter in some models of quantum gravity, by means of identifying appropriate asymmetry parameters in the time evolution of 
intensities (\ref{eq:Intensity:02}) between the two decays in a B factory, based on observables that have already been used in previous studies~\cite{Bernabeu:2016sgz} probing independently T, CP and CPT symmetries in the absence of $\omega$. In the current analysis we have included, simultaneously with the $\omega$, also the conventional CPT parameter $\theta$, already considered in \cite{Bernabeu:2016sgz}, which parameterises CPT violation in the case of a well-defined CPT operator which however does not commute with the hamiltonian of the system, 
indicating a violation of CPT parameterised within the framework of effective field theories (\emph{e.g.} due to Lorentz symmetry violation by a space-time background), in contrast to the parameter $\omega$ that goes beyond that framework.

As we have demonstrated in the present article the set of observables of the B system (\ref{obs3}), (\ref{obs4}), (\ref{obs1}) and (\ref{obs2}) allow for a simultaneous determination (bounds) of the CPT violating parameters $\omega$ and $\theta$, which can thus be disentangled. 
The results obtained from the experimental data from the 
BaBar measurements~\cite{Lees:2012uka}  
(see table \ref{TAB:fit}(I)) are sensitive for the first time to $\im{\om}$,  pointing towards a $2.4\sigma$ deviation from $\im{\om}=0$, which we interpret as an upper bound. 
The observables (\ref{obs1}),(\ref{obs2}) are also sensitive to $\re{\om}$, but they do not show any significant deviation from $\re{\om}=0$, and in this sense they are inferior to 
the previous analyses~\cite{Alvarez:2006ry}  using equal sign semileptonic decay asymmetries of the B system, which yield $\re{\om}=(0.8\pm 4.6)\times 10^{-3}$.
The results (\ref{obs3}) and (\ref{obs4}) also allow a fit to the CPT violating parameter $\theta$, and are compatible with the previous determination in \cite{Bernabeu:2016sgz} and the one performed by the BaBar collaboration in \cite{BABAR:2016tjc}, pointing towards  a $2\sigma$ effect in $\re{\theta}$, also interpreted as an upper bound for the corresponding parameter. 

Moreover, the parameters that measure the presence of wrong flavour decays in $B_d\to J/\Psi K$, i.e. $\C{\KS}-\C{\KL}$, $\S{\KS}+\S{\KL}$ and $\R{\KS}+\R{\KL}$, do not show any significant deviation from zero and the results are consistent with \cite{Bernabeu:2016sgz}.
In the case of $\S{\KS}$ and $\R{\KS}$ we observe that they differ by more than $1\sigma$ with respect to the determination in \cite{Bernabeu:2016sgz} without including the $\omega$ effect. Should this persist in the future, it could affect the precise determination of the unitarity triangle angle $\beta$.

Before closing we stress once more that a quantum-gravity-decoherence-induced CPT violating and entanglement-weakening parameter $\omega$ may not only characterise the initial state of an entangled (neutral) meson system, but may also be generated as a result of a decoherening time evolution that goes beyond the local effective field theory framework~\cite{Bernabeu:2006av}. A full analysis of that case will appear in a forthcoming publication.


 \section*{Acknowledgments\label{SEC:Ack}}
JB, FJB and MN acknowledge financial support from the Spanish MINECO 
 through Grants FPA2015-68318-R, FPA2014-54459-P and the \emph{Severo Ochoa} Excellence Center Project SEV-2014-0398, and from \emph{Generalitat Valenciana} through Grants\\ PROMETEOII/2013/017 and PROMETEOII/2014/049.\newline
The work of NEM is supported in part by the U.K. Science and Technology Facilities Council (STFC) via the grants ST/L000326/1 and ST/P000258/1.\newline
MN acknowledges support from \emph{Funda\c{c}\~ao para a Ci\^encia e a Tecnologia} (FCT, Portugal) through postdoctoral grant SFRH/BPD/112999/2015 and through the project CFTP-FCT Unit 777 (UID/FIS/00777/2013) which are partially funded through POCTI (FEDER), COMPETE, QREN and EU.



%
\clearpage
\providecommand{\href}[2]{#2}\begingroup\raggedright\endgroup


\end{document}